\documentclass[conference]{IEEEtran}

\IEEEoverridecommandlockouts                              

\usepackage{color}
\usepackage{multirow}
\usepackage{cite}
\usepackage{balance}

\usepackage{color}
\usepackage{multirow}
\usepackage{cite}
\usepackage{balance}
\usepackage{amssymb}
\usepackage{cleveref}
\usepackage{soul}

\usepackage[pdftex]{graphicx}
\usepackage[table,xcdraw]{xcolor}
\DeclareGraphicsExtensions{.pdf,.jpeg,.png,.JPG,}
\usepackage{graphicx}

\DeclareGraphicsExtensions{.eps}
\DeclareGraphicsExtensions{.tif}

\graphicspath{{figs/}}
\usepackage{epstopdf}
\usepackage{booktabs}

\usepackage{amsmath}
\usepackage{cleveref}
\ifCLASSOPTIONcompsoc
    \usepackage[caption=false, font=normalsize, labelfont=sf, textfont=sf]{subfig}
\else
\usepackage[caption=false, font=footnotesize]{subfig}
\fi

\begin{document}

\title{Smart Building Energy Management using Nonlinear Economic Model Predictive Control}

\author{\IEEEauthorblockN{Mohammad Ostadijafari and Anamika Dubey}
\IEEEauthorblockA{School of Electrical Engineering and Computer Science\\
Washington State University\\
Pullman, WA}

\and
\IEEEauthorblockN{Yang Liu, Jie Shi, and Nanpeng Yu}
\IEEEauthorblockA{Department of Electrical and Computer Engineering\\
University of California\\
Riverside, CA}}

\maketitle

\begin{abstract}
Owing to the call for energy efficiency, the need to optimize the energy consumption of commercial buildings-- responsible for over 40\% of US energy consumption--has recently gained significant attention. Moreover, the ability to participate in the retail electricity markets through proactive demand-side participation has recently led to development of economic model predictive control (EMPC) for building's Heating, Ventilation,  and  Air  Conditioning  (HVAC) system. The objective of this paper is to develop a price-sensitive operational model for building's HVAC systems while considering inflexible loads and other distributed energy resources (DERs) such  as  photovoltaic  (PV)  generation  and  battery  storage  for the  buildings. A Nonlinear Economic Model  Predictive  Controller (NL-EMPC) is presented  to minimize the net  cost  of  energy  usage by building's  HVAC  system while satisfying the comfort-level of building's occupants. The efficiency of the proposed NL-EMPC controller is evaluated using several simulation case studies.
\end{abstract}

\begin{IEEEkeywords}
Nonlinear economic model predictive control, Building thermal model, HVAC, Demand response, battery, PV.
\end{IEEEkeywords}

\IEEEpeerreviewmaketitle
\section{Introduction}

The Heating, Ventilation, and Air Conditioning (HVAC) system is responsible for a significant proportion of the building total energy consumption. Recently, as a result of wholesale electricity market restructuring and development of retail electricity markets, researchers have explored the potential of commercial buildings in proactive demand-side participation. For example, in \cite{liu2018coordinating}, authors proposed an MPC-based optimization approach to generate proactive demand-bid curves for the buildings to optimally schedule their energy consumption in response to the variable electricity prices. The optimization of the building energy consumption while satisfying the occupants' comfort requirements requires an accurate model for thermal building  loads and advance control methods for the HVAC system. In literature, model predictive control (MPC) for both tracking a desired set-point and for economic optimization (using economic model predictive control/EMPC) has been employed to solve this problem \cite{long2016hierarchical}.

MPC is a model-based controller that requires the dynamical model of the system to obtain optimal control inputs. The required model of the system must be sufficiently accurate to acquire a valid prediction of system states in a computationally tractable manner \cite{cigler2013model}. Building thermal model dynamics and consequently HVAC system model is nonlinear \cite{liu2018coordinating,ma2012demand}. For example, in \cite{ma2012demand}, authors used EnergyPlus, a popular building energy simulation software \cite{crawley2001energyplus}, to simulate thermal and energy behavior of a multi-zone commercial building. However, owing to the high-levels of model complexity, they use a auto-regressive model to approximate the dynamics with a linear dynamical model to develop an EMPC controller for reducing electricity usage cost for building's HVAC system.  
In \cite{maasoumy2012total,haghighi2011modeling,mantovani2015temperature}, Jacobian linearization approach is used to eliminate the system nonlinearity. The resulting linear model is used to design a traditional MPC for temperature set-point tracking. In \cite{rehrl2011temperature}, authors use feedback linearization approach to linearize the simplified nonlinear system model and develop MPC technique to track set-point temperature using water-to-air heat exchange in HVAC systems. By proposing a nonlinear model for the overall cooling system, \cite{ma2012model} presents a MPC scheme for minimizing energy consumption. Assuming the temperature can vary in a short range, \cite{wei2016proactive} propose a MPC-based control algorithm based on Jacobian linearized model to co-schedule the HVAC system control and the battery storage usage for reducing energy cost while meeting HVAC system requirements related to  room temperature set-point and airflow. 

	
Unfortunately,  the  Jacobian  linearization  approach  is  not valid  when the  desired  room  temperature  obtained  from  the optimization problem problem vary significantly at different time-steps. This is usually the case when the building is not occupied for certain time of the day and can be overheated or overcooled to  achieve  the  desired  economic  objective.  This  case  is  of significant  interest  when  optimizing the transacted  cost of  energy  by  leveraging  the  occupancy  information  of  the building. Since, the primary energy consumption for a building is due to its HVAC system, significant energy savings can be achieved using a price-sensitive HVAC model that optimally schedules heating/cooling while taking the building's occupancy information into account, as demonstrated using a nonlinear MPC-based optimization problem in \cite{liu2018coordinating}. However, they do not consider the co-scheduling of other energy resources using a price-sensitive model. 



The objective of this paper is to develop a price-sensitive operational model for building’s HVAC systems while considering inflexible loads and other distributed energy resources (DERs) such as photovoltaic (PV) generation and battery storage. Note that in the proactive setting, an efficient HVAC controller should track an optimal temperature trajectory based on HVAC dynamical model, comfort ranges based on occupancy information, and weather forecasts while taking time-dependent cost of energy into account \cite{parisio2013scenario}. Inspired by \cite{liu2018coordinating} and \cite{wei2016proactive}, in this work, we present a Nonlinear Economic Model Predictive Controller (NL-EMPC) that minimizes the net cost of energy usage by the HVAC system with an {\em{imperfect prediction of the future disturbance vectors}}. In addition, we address the problem of {\em {co-scheduling HVAC with other DERs and inflexible loads of the building}} using the proposed NL-EMPC and demonstrate the added cost savings. 


\section{Overview of the Proposed Framework}
\label{Overview of the proposed framework}
This section details building thermal load, battery energy storage, and PV panel. 

\vspace{-0.1cm}
\subsection{Building Thermal Model}
	\label{building thermal model}
	Thermal model of a building is usually obtained by modeling the building as a first-order RC network  \cite{haghighi2011modeling,maasoumy2012total}. In the resulting RC network, a node indicates a wall or a room. In general, if there are in total $n$ nodes, $m$ of which denote rooms, then, $n-m$ remaining nodes denote walls. Using the same equations as detailed in \cite{haghighi2011modeling,maasoumy2012total} to describe rooms and walls temperatures, and after zero-hold discretization \cite{liu2018coordinating}, we obtain the following state-space equations representing the building thermal model:
	\begin{equation}\label{eq3}
	\small
	\boldsymbol{x}^{k+1}=\boldsymbol{A}\boldsymbol{x}^{k}+\boldsymbol{B}\boldsymbol{u}^{k}\circ(\boldsymbol{T_s}-\boldsymbol{y}^{k})+\boldsymbol{E}\boldsymbol{d}^{k}
	\end{equation}
	\begin{equation}\label{eq4}
	\small
    \boldsymbol{y}^{k}=\boldsymbol{C}\boldsymbol{x}^{k}
	\end{equation} 
	where superscript $k$ shows the sampling time and $\circ$ is the element-wise product operator for two vectors; $\boldsymbol{d}^{k} \in \mathbb{R}^{l}$ is the vector of disturbance (with $l$ number of the disturbance elements such as external temperature, solar radiation and internal gains, etc.) at sampling time $k$; $\boldsymbol{A} \in \mathbb{R}^{n\times n}$, $\boldsymbol{B} \in \mathbb{R}^{n\times m}$, $\boldsymbol{C} \in \mathbb{R}^{m\times n}$ and $\boldsymbol{E} \in \mathbb{R}^{n\times l}$ are matrices obtained from building thermal model representing time-invariant building parameters (see \cite{maasoumy2012total,haghighi2011modeling} for more details); $\boldsymbol{x}^{k} \in \mathbb{R}^{n}$ is the state vector representing the temperature of the network nodes; $\boldsymbol{u}^k \in \mathbb{R}^{m}$ is the vector of input variables whose elements ($u^{k}_i$) are mass flow into each thermal zone; $\boldsymbol{y}^{k} \in \mathbb{R}^{m}$ is the output vector of the system; and $\boldsymbol{T_s}\in \mathbb{R}^{m}$ with entries of $T_{s_i}$ representing the temperature of supply air to the room $i$.
	
	In \cite{liu2018coordinating}, it is assumed that the perfect prediction of disturbance input over the prediction window is available at each sampling time. 
	A perfect prediction of the future disturbance is not possible in practice due to the uncertainty in forecasting ambient temperature, solar radiation and other weather-related effects\cite{mantovani2015temperature}. The dynamics of the system after considering prediction error and the new state-space representation for the thermal building model is detailed in (\ref{eq5}) and (\ref{eq6}).
	    \begin{equation}\label{eq5}
    \small
	\boldsymbol{D}^k=\boldsymbol{d}^{k}+\boldsymbol{\epsilon}^{k}
	\end{equation}
	\begin{equation}\label{eq6}
	\small
	\boldsymbol{x}^{k+1}=\boldsymbol{A}\boldsymbol{x}^{k}+\boldsymbol{B}\boldsymbol{u}^{k}\circ(\boldsymbol{T_s}-\boldsymbol{y}^{k})+\boldsymbol{E}\boldsymbol{D}^{k}
	\end{equation}
	
	\noindent where $\boldsymbol{\epsilon}^{k}$ is error due to uncertainty in predicting $\boldsymbol{d}^{k}$. 
	

	Next, we detail the equations for power consumption of HVAC system as a function of mass flow rate ($u_i^k$). A typical HVAC system consumes most of its power through heater, chiller, and fan. Without loss of generality, in this paper we only consider a cooling system. The fan power consumption, $P^{k}_{f_i}$, is modeled as a cubic function of air mass flow rate \cite{liu2018coordinating}:
	\begin{equation}\label{eq7}
	P^{k}_{f_i}=P_{rated_i} (u^{k}_i/u_{rated_i})^3 
	\end{equation}
	where $P_{rated_i}$ and $u_{rated_i}$ are the rated power and rated outlet mass flow rate of the air handling unit of HVAC system in thermal zone $i$, respectively; and $P_{f_i}^k$, $u^{k}_i$ are power consumption and the air mass flow rate (control variable) of fans in thermal zone $i$ at sampling time $k$, respectively. 
	
	The cooling load is a function of the air mass flow rate, ambient temperature, and temperature of the thermal zone $i$  as defined in \cite{liu2018coordinating}:
		\vspace{-0.2cm}
	\begin{equation}\label{eq8}
	\small
	P^{k}_c=\frac{c_a}{COP}\sum^{m}_{i=1}{u^{k}_i\left[d_p y^k_i+(1-d_p)T^k_{out}-T_{s_i}\right]} 
	\vspace{-0.2cm}
	\end{equation}
	where $T^k_{out}$ is the ambient temperature at sampling time $k$; $COP$ is performance coefficient of the chiller; $c_a$ is the specific heat capacity of the air; and $d_p$ is the instantaneous return-to-total ratio of the chiller that varies between $0$ and $1$. 
	
	Therefore, the total power consumption of the entire building by its HVAC  system at sampling time $k$ is given by (\ref{eq9}).
	\vspace{-0.3cm}
	\begin{equation}\label{eq9}
	\small
	 P^{k}_H=P^{k}_c+\sum^{m}_{i=1}P^{k}_{f_i}
	 \vspace{-0.14cm}
	\end{equation}

\vspace{-0.3cm}
\subsection{Battery Energy Storage Model}
\label{Battery storage model}
The dynamics for battery energy storage can be formulated using the state-space equations for the state-of-charge (SOC) and the limits on battery's charging and discharging power and energy as follows: 
\vspace{-0.35cm}
    \begin{equation}\label{eq10}
    \small
	{SOC}^{k+1}=(1-\nu)SOC^{k}+\rho \frac{P^{k}_{c,d}}{Q_{bat}}\tau
	\end{equation}
	\vspace{-0.4cm}
	\begin{equation}\label{eq11}
	\small
	E^{-}\leqslant {SOC}^{k+1} \leqslant E^{+}
	\end{equation}
	\vspace{-0.4cm}
	\begin{equation}\label{eq12}
	\small
	-d_r\leqslant P^{k}_{c,d} \leqslant c_r 
	\end{equation}
Specifically, the battery SOC updates based on (\ref{eq10}) where $SOC^{k}$ and $P^{k}_{c,d}$ are the SOC and charging/discharging power of battery at sampling time $k$, respectively; $\nu$, $\rho$, $Q_{bat}$ and $\tau$ are energy decay rate, round-trip efficiency, capacity of the battery and length of time step respectively. Constraint (\ref{eq11}) guarantees the battery SOC remains in safety boundary where, $E^{+}$ and $E^{-}$ specifies bounds on battery charging/discharging limits. Finally, constraint (\ref{eq12}) bounds batteries maximum charging/discharging rates where $d_r$ is the maximum discharge rate while $c_r$ is the maximum charge rate. It should be noted that in the above formulation when $P^{k}_{c,d}>0$, the battery is charging and when $P^{k}_{c,d}<0$, the battery is discharging. 
\subsection{Photovoltaic (PV) Generator Model}
The PV panel is modeled as a negative load with rated active power of $P_{PV}^{rated}$ and an associated multiplier $\alpha^k$ indicating the effect of variation in solar irradiance at the sampling time $k$. The PV generation at sampling time $k$, $P_{PV}^k$, is given by (\ref{eqPV}).

\vspace{-0.3cm}
\begin{equation}\label{eqPV}
\small
P_{PV}^k = 	\alpha^k. P_{PV}^{rated}
\vspace{-0.3cm}
\end{equation}
	
	  	\begin{figure}[t]
		\centering
			\vspace{-0.5cm}
		\includegraphics[width=6cm,height=4cm]{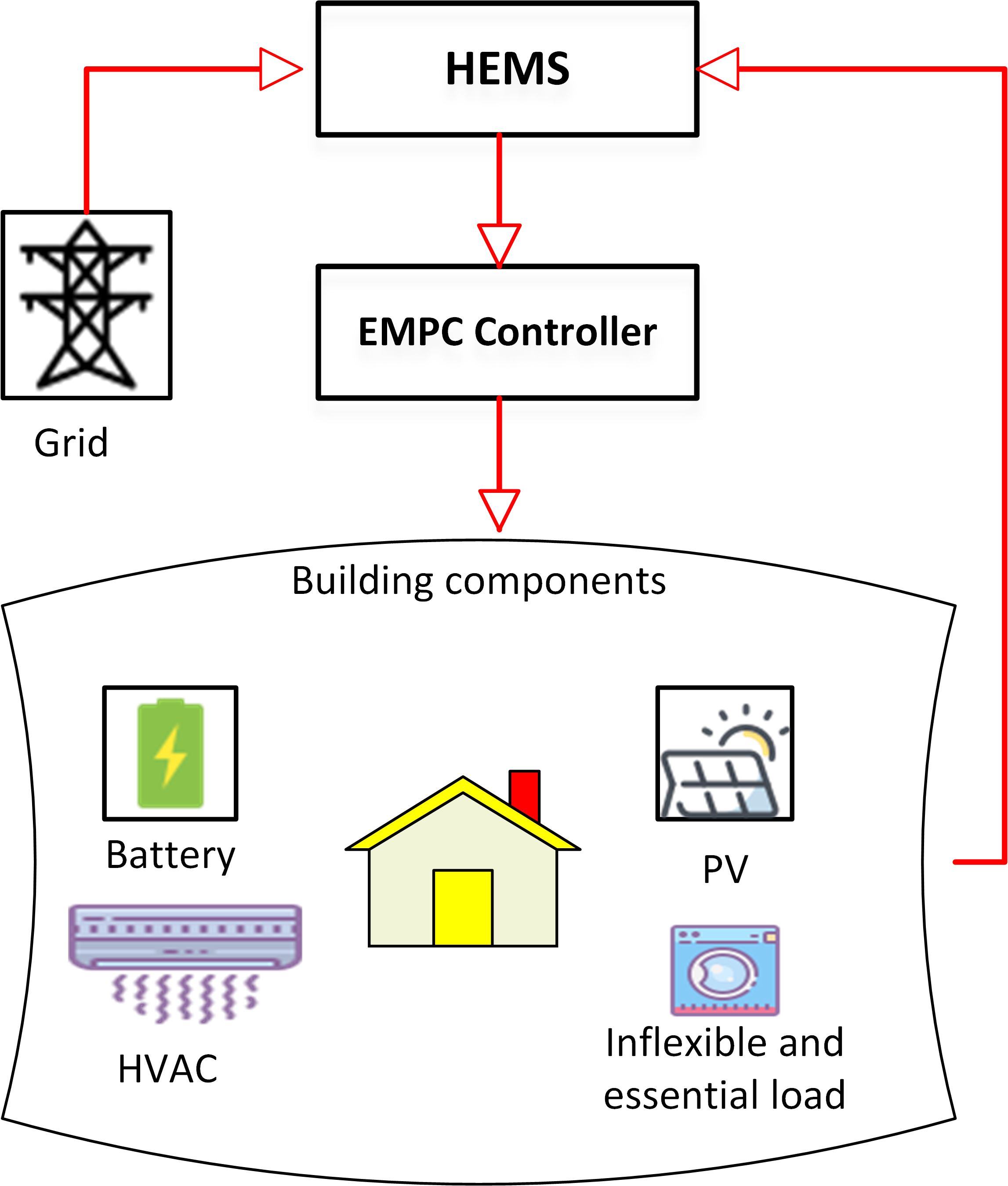}
		\vspace{-0.2cm}
		\caption{System layout}
		\label{fig:1}
		\vspace{-0.5cm}
	\end{figure}
	
\section{Optimal Scheduling of Power Consumption}
\label{Optimal Scheduling of energy}
	Fig. \ref{fig:1} shows the component layout used in this paper. It is assumed that the building is equipped with the home energy management system (HEMS) \cite{shareef2018review}. In order to satisfy the building's energy demand, HEMS can provide electricity from any combination of PV generation, battery storage, and electricity purchased from the retail electricity provider \cite{cai2018economic}. 
	
	The objective in this section is to optimally co-schedule the HVAC system with inflexible loads and available energy resources of the building such that it can optimize the net cost of transacted energy for the specified prediction window while ensuring that the desired level of comfort is met for its occupants. The problem is formulated as an economic model predictive control (EMPC) problem with the objective of minimizing the building's total electricity usage for a given price vector, whose entries indicate time-of-use (TOU) electricity tariffs for each hour of the day subject to the thermal building model detailed in (\ref{eq3})-(\ref{eq4}) and (\ref{eq7})-(\ref{eq18}). 
	
	\vspace{-0.1cm}
	\begin{small}
    \begin{equation}\label{eq13}
	\underset{\boldsymbol{u}^{k}}{Min}\sum^{t+W-1}_{k=t}{{Price}^{k}.P^{k}_{T}}
	\end{equation}
	\end{small}
	Subject to:
    \vspace{-0.1cm}
	\begin{small}
    \begin{equation}\label{eq14}
	T^{k}_{Min}\leqslant T^{k}_r \leqslant T^{k}_{Max}
	\end{equation}
	\end{small}
	    \vspace{-0.3cm}
	\begin{small}
    \begin{equation}\label{eq15}
	0\leqslant P^{k}_H \leqslant P^k_{H_{Max}}
	\end{equation}
	\end{small}
    \vspace{-0.3cm}
	\begin{small}
    \begin{equation}\label{eq16}
	u^{k}_{Min}\leqslant u^{k} \leqslant u^{k}_{Max}
	\end{equation}
    \vspace{-0.3cm}
	\begin{small}
    \begin{equation}\label{eq17}
	P^{k}_T=P^{k}_H+P^{k}_l+P^{k}_{c,d}-P^{k}_{pv}
	\end{equation}
	\end{small}
	\vspace{-0.3cm}
	\begin{small}
    \begin{equation}\label{eq18}
	P^{k}_T\geqslant 0
	\end{equation}
	\end{small}
	\vspace{-0.5cm}
	\begin{align*}
	\text{and constraints (\ref{eq3})-(\ref{eq4}) and (\ref{eq7})-(\ref{eqPV}) }
	\end{align*}
	\end{small}
	The minimization of the electricity usage cost is given by (\ref{eq13}), where $P^{k}_T$ is the electric power purchased from the retail electricity provider; $W$ is the prediction window, and $Price^k$ is the electricity tariff at the sampling time $k$. The desired temperature range, HVAC power consumption limits, air mass flow limits, the thermal building model, and the total consumed power by HVAC are presented in (\ref{eq14}), (\ref{eq15}), (\ref{eq16}), (\ref{eq3})-(\ref{eq4}), and (\ref{eq7})-(\ref{eq9}), respectively; where, at sampling time $k$, variables $T^{k}_{Min}, T^{k}_{Max}, P^{k}_{H_{Max}}, u^{k}_{Min}$ and $u^{k}_{Max}$ are minimum and maximum range of the temperature ($^{\circ}C$), maximum HVAC power consumption limits, minimum and maximum limits for HVAC mass flow rate, respectively. Constraint (\ref{eq17}) determines the total power consumption where $P^k_l$ is the consumed power by inflexible loads of the building at sampling time $k$. Constraint (\ref{eq18}) states that the total power consumption cannot be negative, in other words, the surplus of energy cannot be sold back to the power grid. 
	
	It should be noted that (\ref{eq3}) is bilinear in system input and states which results in a nonlinear economic model-predictive control (NL-EMPC) problem. Note that Jacobian-linearization methods, extensively used in related literature for temperature set-point tracking, is not a valid approach to solve the aforementioned problem. That is, Jacobian-linearization is not valid when a wide-range of temperature variations in buildings are expected due to varying occupancy patterns when attempting to optimize electricity usage given time-varying cost of electricity. Therefore, a fully nonlinear model needs to be solved for the case under consideration.  
	
	
	\begin{figure}[t]
		\centering
		\vspace{-0.2cm}
		\includegraphics[width=8cm, trim={0 0 0 0 },clip]{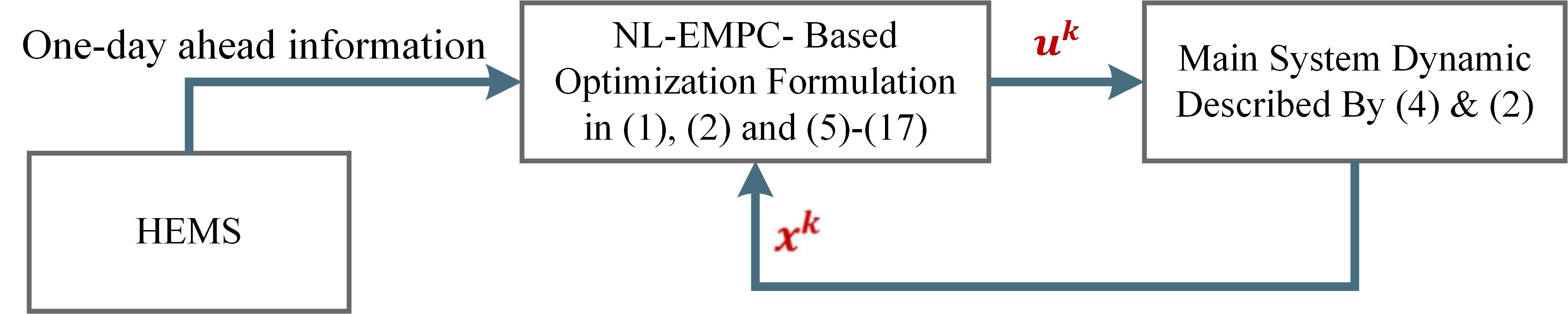}
		\vspace{-0.4cm}
		\caption{Diagram of the NL-EMPC controller}
		\label{fig:2}
		\vspace{-0.6cm}
	\end{figure}

A schematic view of the proposed NL-EMPC is detailed in Fig. \ref{fig:2}. At the beginning of each day, HEMS provides NL-EMPC controller, one-day ahead prediction information including the  occupancy pattern, PV generation, power consumption by inflexible loads, and TOU prices for the next 24 hours of the day. This information determines $T^{k}_{Min}$ and $T^{k}_{Max}$ in (\ref{eq14}), $P^{k}_{pv}$ and $P^{k}_{c,d}$ in (\ref{eq17}), and $Price^k$ in objective function (\ref{eq13}). NL-EMPC algorithm solves minimization problem (\ref{eq13}), with constraints  (\ref{eq3}), (\ref{eq4}) and (\ref{eq7})-(\ref{eq18}) at each sampling time $k$. This results in optimal mass air flow rate trajectory [$\boldsymbol{u}^t,\boldsymbol{u}^{t+1}, ...,\boldsymbol{u}^{t+W-1}$] for a prediction window from time $t$ to time $t+W-1$. After obtaining the optimal mass air flow rate trajectory, only the first control input ($\boldsymbol{u}^t$) is applied to the main system plant which is governed by (\ref{eq6}) and (\ref{eq4}). After observing the new values of the system states ($\boldsymbol{x}^{t+1}$), the NL-EMPC algorithm moves one step forward. Using the observed system states as the new initial condition, the minimization problem is solved again from time intervals $t+1$ to $t+W$. The same process continues for the next time steps, and repeatedly a constrained optimization problem over a moving time horizon is solved to choose the control actions using predictions of future costs, disturbances, and constraints. This control method is also known as \textit{receding horizon control approach}. Note that although in this work, it is assumed that NL-EMPC has the knowledge of the exact value of PV generation and power consumption of inflexible loads during the day, the uncertainty in these variables can be treated same as uncertainty in $d^k$ as illustrated in (\ref{eq5}) and (\ref{eq6}). Also, it should be noted that for the case that some or all the states cannot be measured, an observer should be designed to predict the state variables \cite{zhou2014multivariable}. The design of the observer is, however, outside of the scope of this paper.
\vspace{-0.2cm}
\section{Simulation Results}
\label{Simulation Results}
\vspace{-0.1cm}
In this section, we conduct a set of experiments to validate the efficiency of the proposed NL-EMPC controller. 
For thermal building model, we consider a thermal zone with 7 states (four states for temperature of walls, two states for temperature of floor and ceiling, and one state for indoor thermal zone temperature) with the parameters as same as \cite{liu2018coordinating,haghighi2011modeling}. Other building parameters are:  $d_p=0$ , $P_{rated}$ and $u_{rated}$ are $600 W$ and $1 kg/s$, respectively. And battery capacity $Q_{bat}=6 kWh$. 

\begin{figure}[!t]
     \centering
    \subfloat[Ambient Temperature\label{fig:4}]{
		\includegraphics[width=0.45\linewidth]{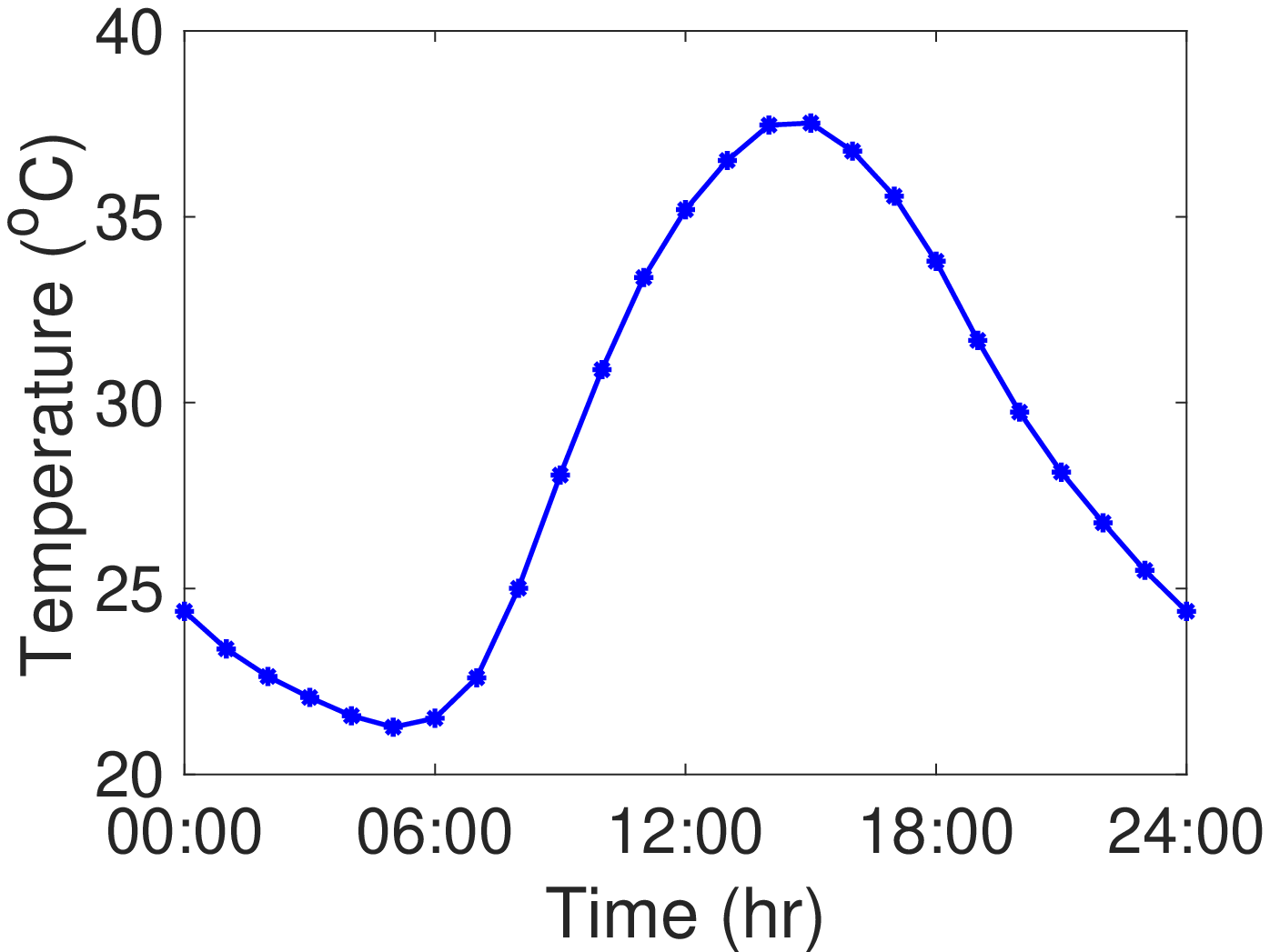}}
		\hfill
	    \subfloat[TOU electricity tariffs\label{fig:5}]{
		\includegraphics[width=0.45\linewidth]{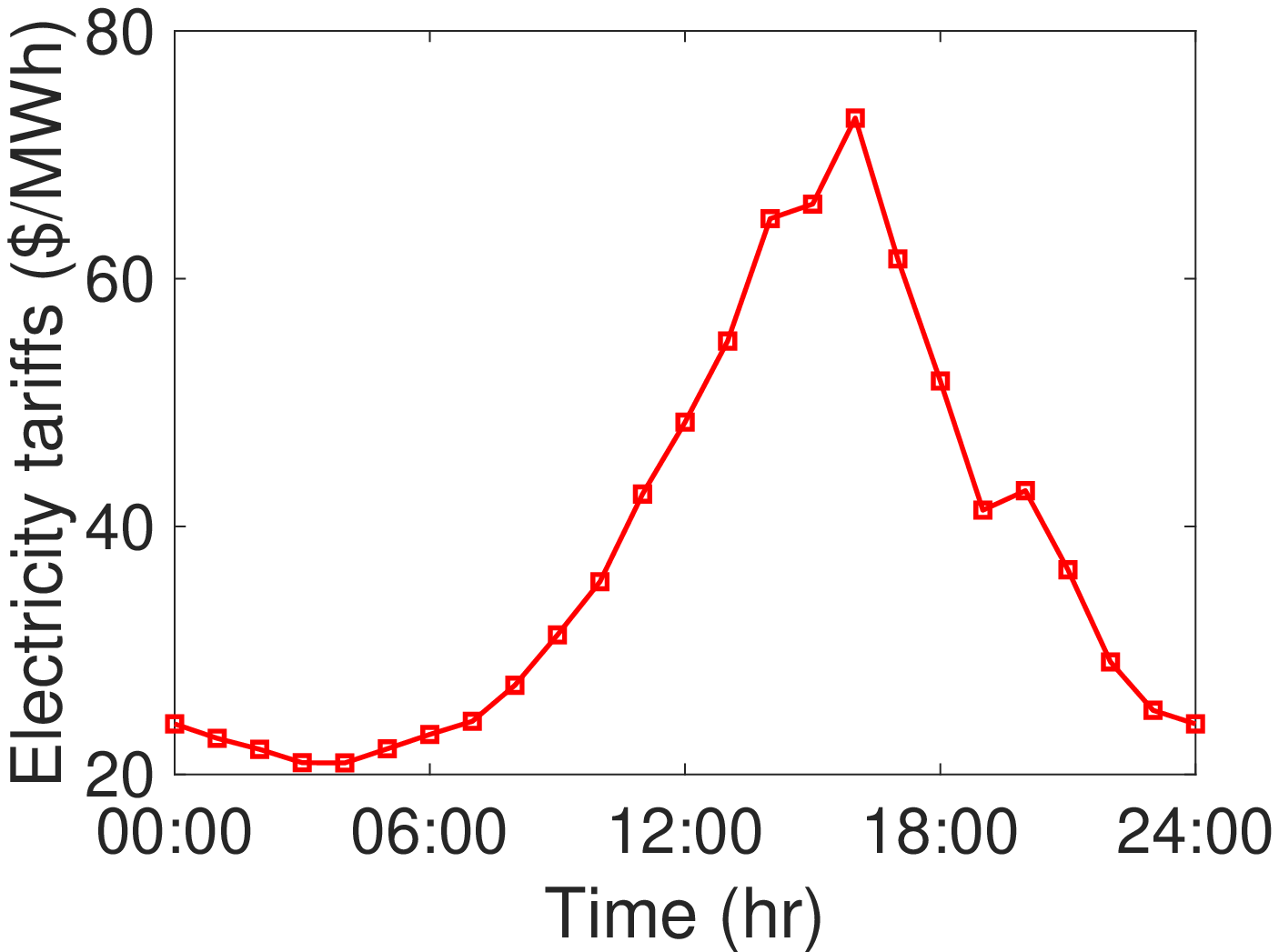}}
		
\subfloat[Received occupancy patterns at the beginning of the day\label{fig:213}]{
\centering
\resizebox{8.5cm}{!}{%
\begin{tabular}{llllllllllllll}
\cline{1-13}
\multicolumn{1}{l|}{Occupancy pattern 1} & \multicolumn{1}{l|}{\cellcolor[HTML]{000000}{\color[HTML]{000000} }} & \multicolumn{1}{l|}{\cellcolor[HTML]{000000}{\color[HTML]{000000} }} & \cellcolor[HTML]{000000}{\color[HTML]{000000} } & \multicolumn{1}{l|}{} & \multicolumn{1}{l|}{} & \multicolumn{1}{l|}{} & \multicolumn{1}{l|}{} & \multicolumn{1}{l|}{} & \multicolumn{1}{l|}{} & \multicolumn{1}{l|}{\cellcolor[HTML]{000000}} & \multicolumn{1}{l|}{\cellcolor[HTML]{000000}} & \multicolumn{1}{l|}{\cellcolor[HTML]{000000}} &  \\ \cline{1-4} \cline{6-13}
\multicolumn{1}{l|}{Occupancy pattern 2} & \multicolumn{1}{l|}{} & \multicolumn{1}{l|}{} & \multicolumn{1}{l|}{} & \cellcolor[HTML]{000000}{\color[HTML]{000000} } & \cellcolor[HTML]{000000}{\color[HTML]{000000} } & \cellcolor[HTML]{000000}{\color[HTML]{000000} } & \cellcolor[HTML]{000000} & \cellcolor[HTML]{000000} & \cellcolor[HTML]{000000} & \multicolumn{1}{l|}{} & \multicolumn{1}{l|}{} & \multicolumn{1}{l|}{} &  \\ \cline{1-13}
\multicolumn{2}{r}{00:00} &  & \multicolumn{2}{l}{06:00} &  & \multicolumn{2}{l}{12:00} &  & \multicolumn{2}{l}{18:00} & \multicolumn{1}{r}{} & \multicolumn{2}{l}{24:00} \\
 &  &  &  & \multicolumn{3}{l}{Time (hr)} &  &  &  &  &  &  & 
\end{tabular}%
}}
\caption{Model Parameters of NL-EMPC for HVAC system}
\vspace{-0.7 cm}
\end{figure}

The predicted ambient temperature received at the beginning of the day is shown in Fig.\ref{fig:4}. The 24-hour TOU electricity tariffs are shown in Fig. \ref{fig:5}. Two occupancy patterns are considered for the building in simulations (see Fig. \ref{fig:213}). To maintain the desired comfort level of building occupants, it is assumed that during occupancy, the indoor temperature in thermal zone should lie between 21- 25 ($^{\circ}C$), otherwise, there is no limit for the thermal zone temperatures. There is no temperature limits for other 6 states of the thermal zones for all the times. The simulations are carried out using Ipopt slover integrated with MATLAB using Opti toolbox. Ipopt is a software package suitable for solving large-scale nonlinear optimization problems. The initial conditions for solving the nonlinear optimization problem are randomly generated to satisfy the prespecified upper and lower bounds for the variables. 


%
%
%
%
 \vspace{-0.6 cm}
 \subsection{Effectiveness of NL-EMPC for HVAC Control}
\begin{figure}[t]
     \centering
    \subfloat[Occupancy pattern 1\label{T1}]{
		\includegraphics[width=0.45\linewidth]{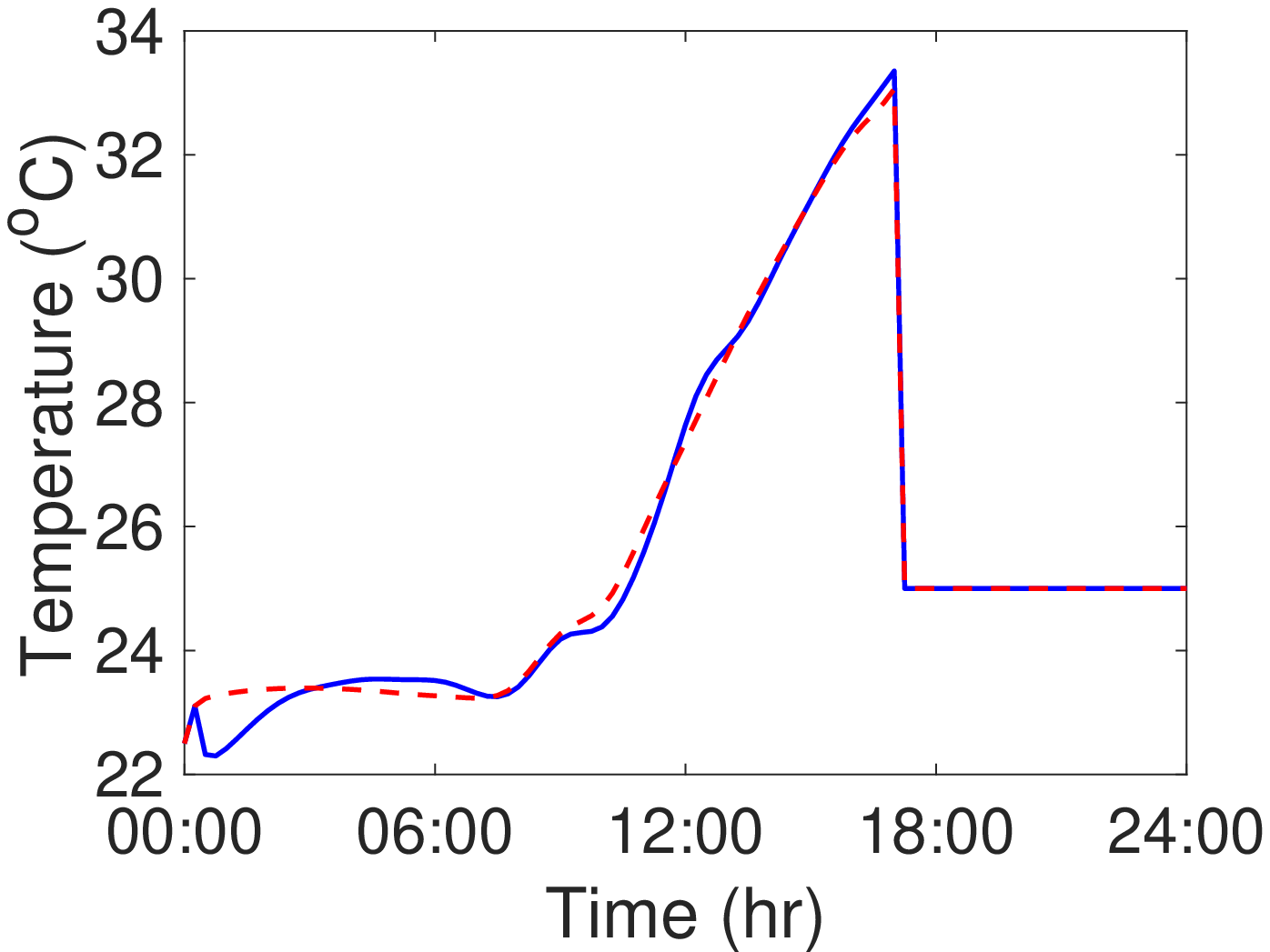}}
		\hfill
	    \subfloat[Occupancy pattern 2\label{T2}]{
		\includegraphics[width=0.45\linewidth]{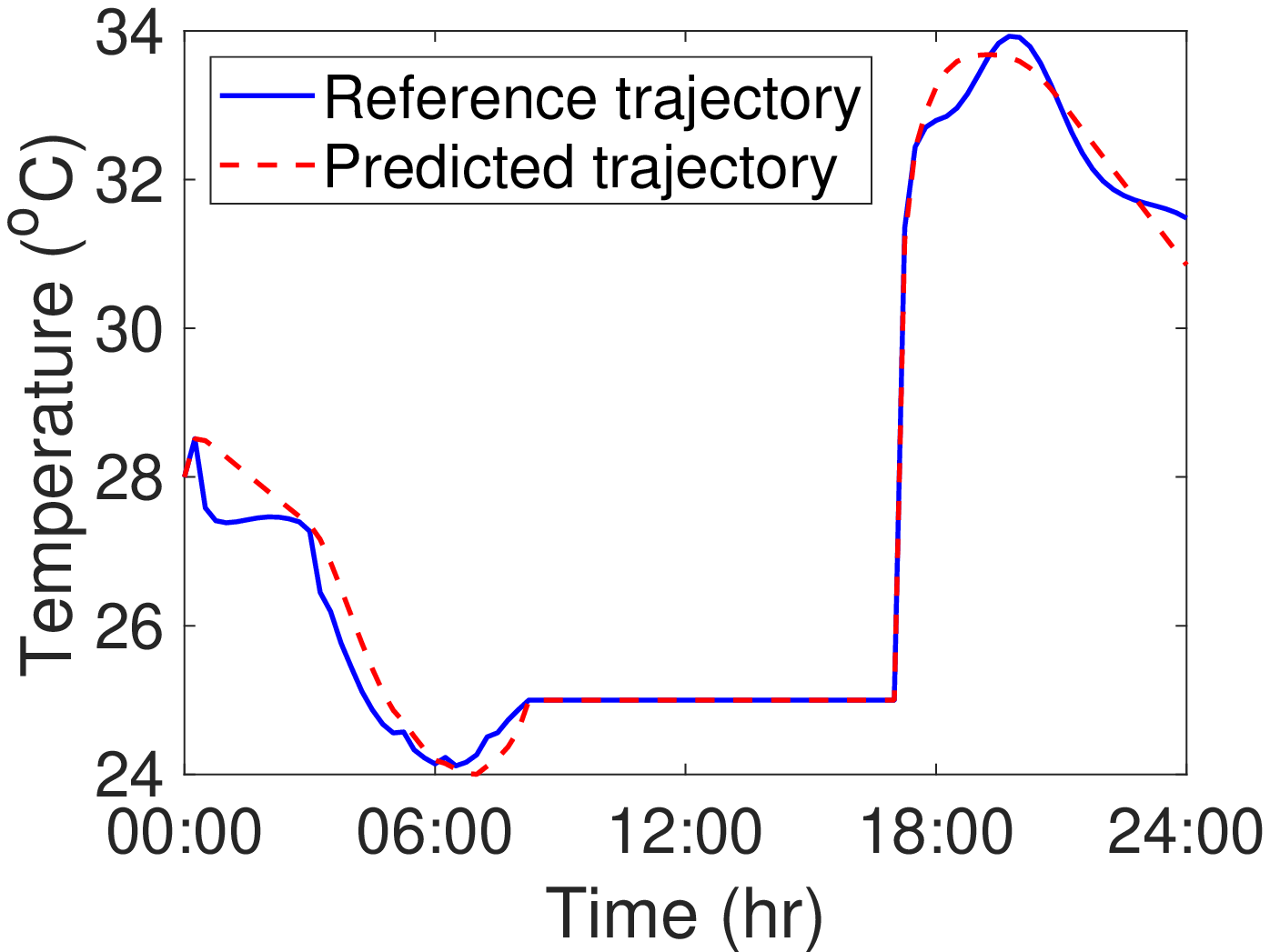}}
  \caption{Indoor thermal zone temperature: reference  trajectory  obtained  from  receding  horizon  approach vs. predicted trajectory at first sampling time.}
\label{fig:9}
\vspace{-0.8 cm}
\end{figure}

 This section validates the proposed NL-EMPC controller for its effectiveness in minimizing the transacted energy cost while accounting for the error in disturbance vector prediction. Starting at the beginning of a day (00:00) and after receiving one-day ahead information, the controller solves the NL-EMPC optimization problem for the next 24 hours at sampling rate of 15 minutes.  We model the error in disturbance vector, $\boldsymbol{\epsilon}^{k}$, as zero mean Gaussian noise with variance of $2$.
 For both occupancy patterns, Fig. \ref{fig:9} shows the evolution of indoor thermal zone temperature obtained using receding horizon approach (reference trajectory) and using optimal controls obtained by solving NL-EMPC at sampling time $k=1$ (predicted trajectory). Note that the receding horizon control approach generates reference trajectory by repeatedly solving NL-EMPC problem at each sampling time and applying the optimal air mass flow control input to HVAC system. On the other hand, the predicted trajectory is obtained using the optimal air mass flow calculated at the first sampling time for all 24 hours. As expected, there are deviations between the predicted and receding horizon trajectories. Notice that the predicted trajectory is obtained with the assumption of having prefect knowledge of the future input disturbances. Thus, the uncertainty in prediction of ($\boldsymbol{\epsilon}^{k}$), is ignored when solving for predicted trajectory. However, the receding horizon approach is able to take this uncertainty into account when re-optimizing the problem at each time step. This case study highlights the role of MPC on managing uncertainties in the dynamical model of HVAC system.
 
\begin{figure}[t]
     \centering
    \subfloat[Occupancy pattern 1\label{P1}]{
		\includegraphics[width=0.45\linewidth]{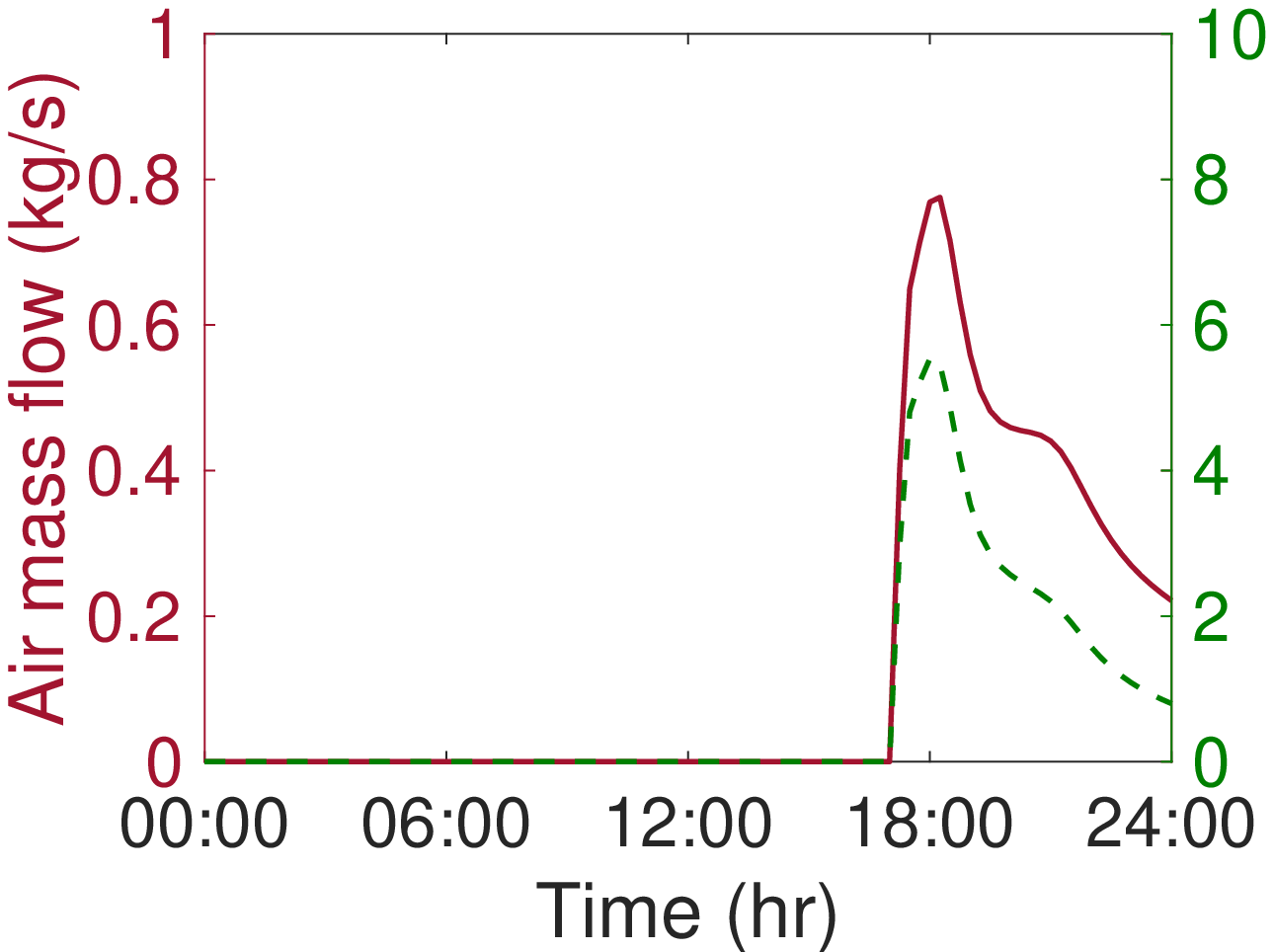}}
	    \hfill
	\subfloat[Occupancy pattern 2\label{P2}]{
		\includegraphics[width=0.45\linewidth]{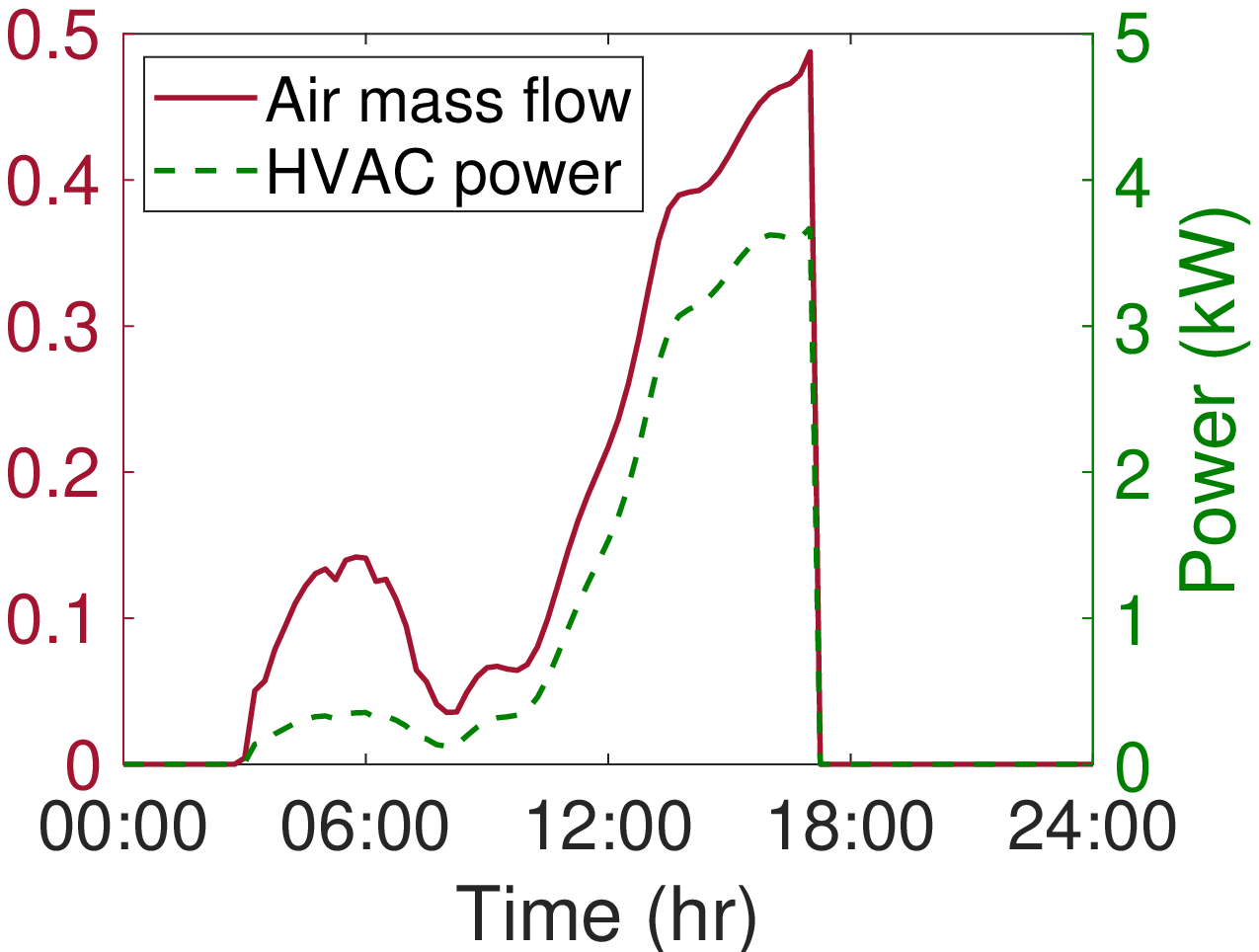}}
  \caption{Total HVAC power consumption and Air mass flow rate calculated by proposed controller based on occupancy patterns.}
\label{fig:10}
\vspace{-0.5 cm}
\end{figure}
 
 Next, Fig. \ref{fig:10} shows the optimal value of the control input $u$ (air mass flow rate) and HVAC power consumption for the both occupancy patterns. As it can be seen in Fig. \ref{fig:9} and Fig. \ref{fig:10}, when the thermal zone (building) is occupied, the controller adjusts the control variable, $u$, of HVAC cooling system such that the temperature of the thermal zone lie within the prespecified comfort range while simultaneously minimizing the cost of transacted energy. On the contrary, when, there is no occupancy in the thermal zone, controllers minimize the total cost of using energy by turning the HVAC cooling system off. Note that there are times during the day (e.g. 00:00-06:00 for occupancy pattern 1) that although the thermal zone is occupied, there is no need to turn HVAC on ($u=P_H=0$). That is, the ambient temperature at these times are low and sufficient to maintain the thermal-zone temperature within the occupants' comfort level without requiring HVAC cooling system. 
 
 
 The price-sensitivity of the model is emphasized for the optimal controls obtained for occupancy pattern 2. Notice that although the building is unoccupied till 6:00, the HVAC control is ON from 4:00-6:00. Due to low TOU electricity tarrifs, the optimal solution is to precool the building from 4:00-6:00 by turning ON HVAC, so as to consume a smaller amount of expensive electricity after 6:00. The NL-EMPC controller thus leverages the thermal building  dynamics to minimize the overall cost of transacted energy.

	\begin{figure}[t]
     \centering
     \vspace{-0.5cm}
    \subfloat[PV generation vs irradiation \label{fig:7}]{
		\includegraphics[width=0.48\linewidth]{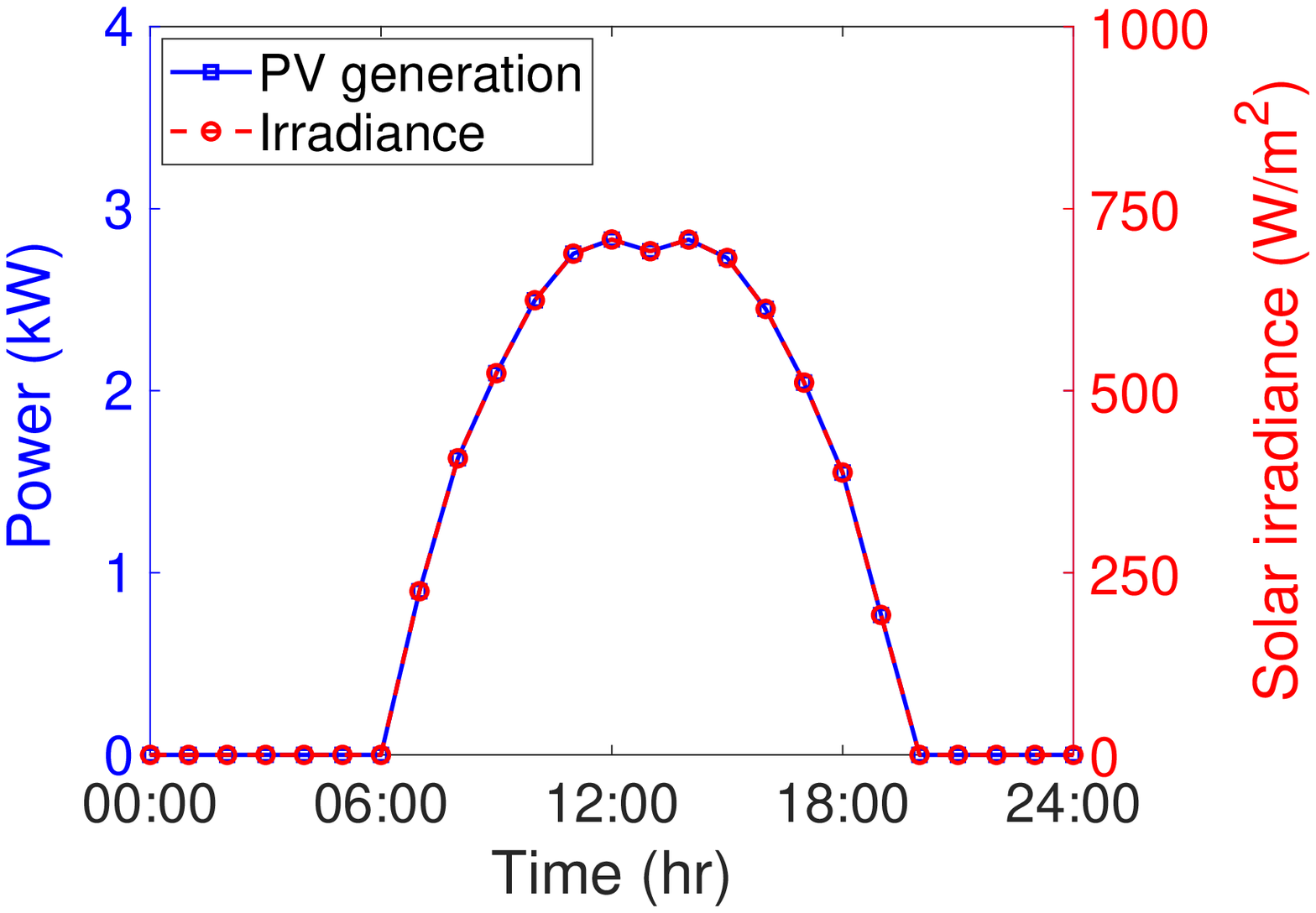}}
	    \hfill
	\subfloat[Power demand of inflexible load \label{fig:8}]{
		\includegraphics[width=0.45\linewidth]{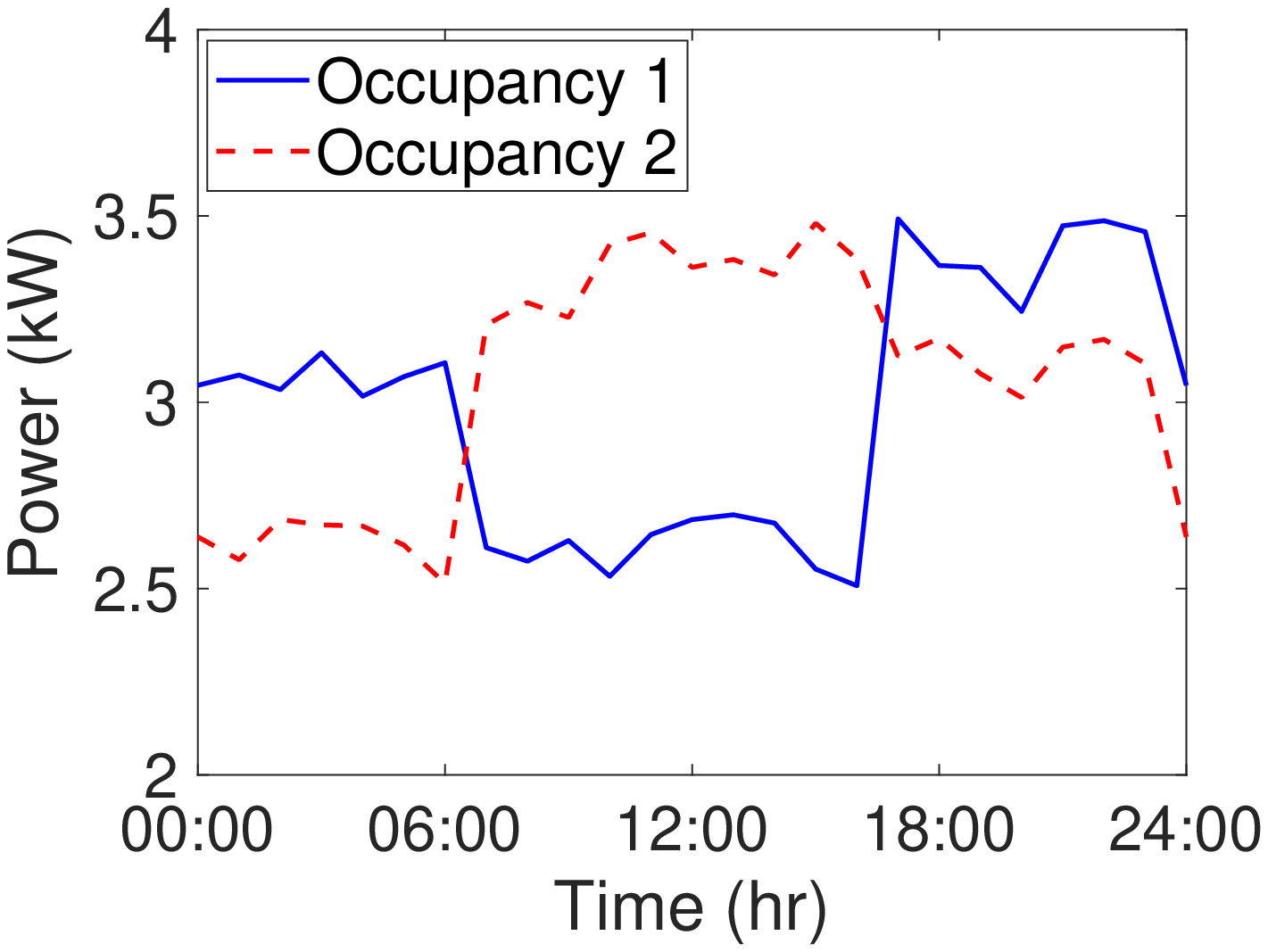}}
  \vspace{-0.4 cm}
  \caption{Model Parameter for Co-Scheduling Energy}
\label{fig:new}
\vspace{-0.6 cm}
\end{figure}

	\begin{figure}[t]
     \centering
    \subfloat[Occupancy pattern 1 \label{aaa}]{
		\includegraphics[width=0.46\linewidth]{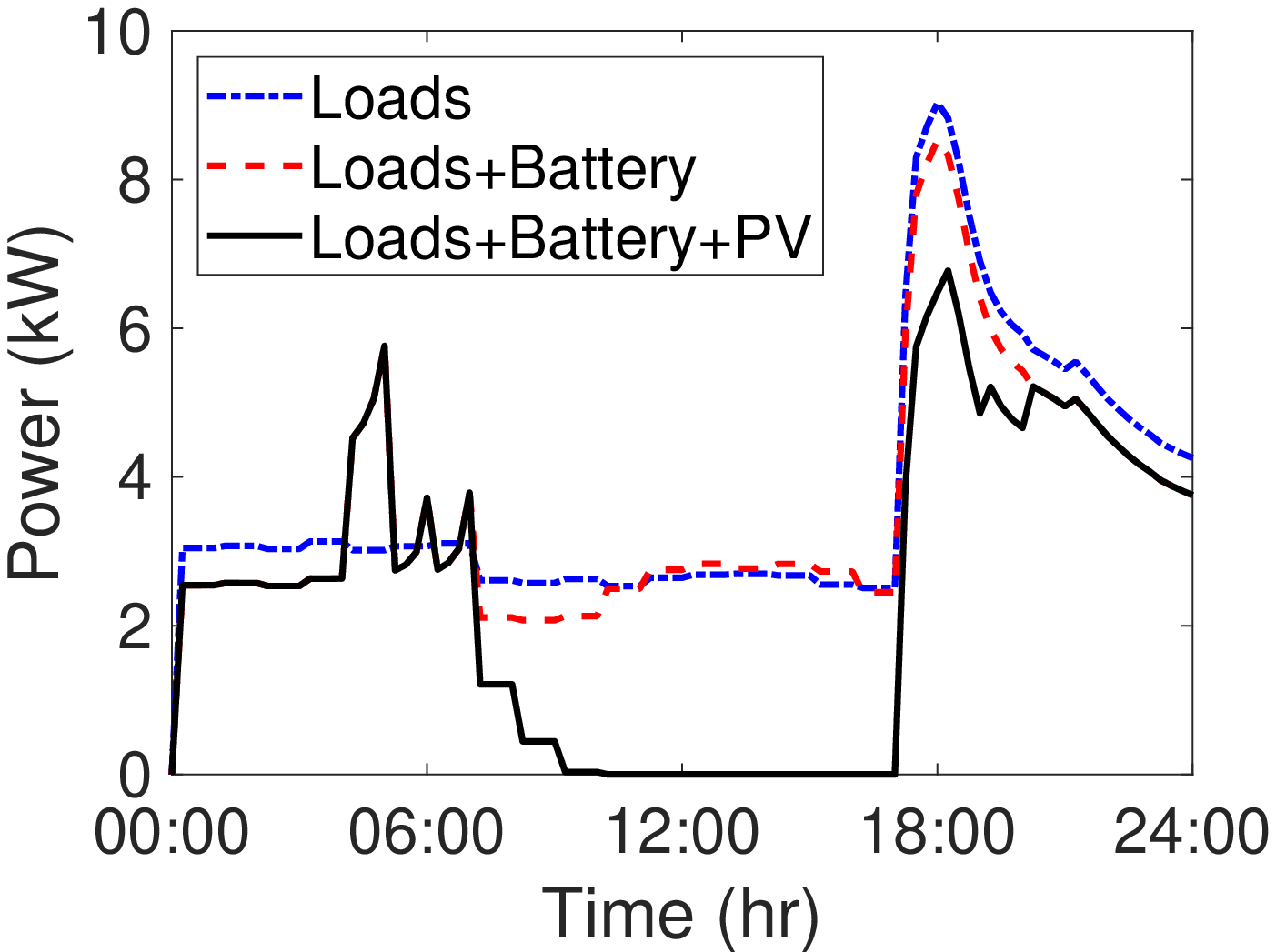}}
	    \hfill
	\subfloat[Occupancy pattern 2\label{bbb}]{
		\includegraphics[width=0.46\linewidth]{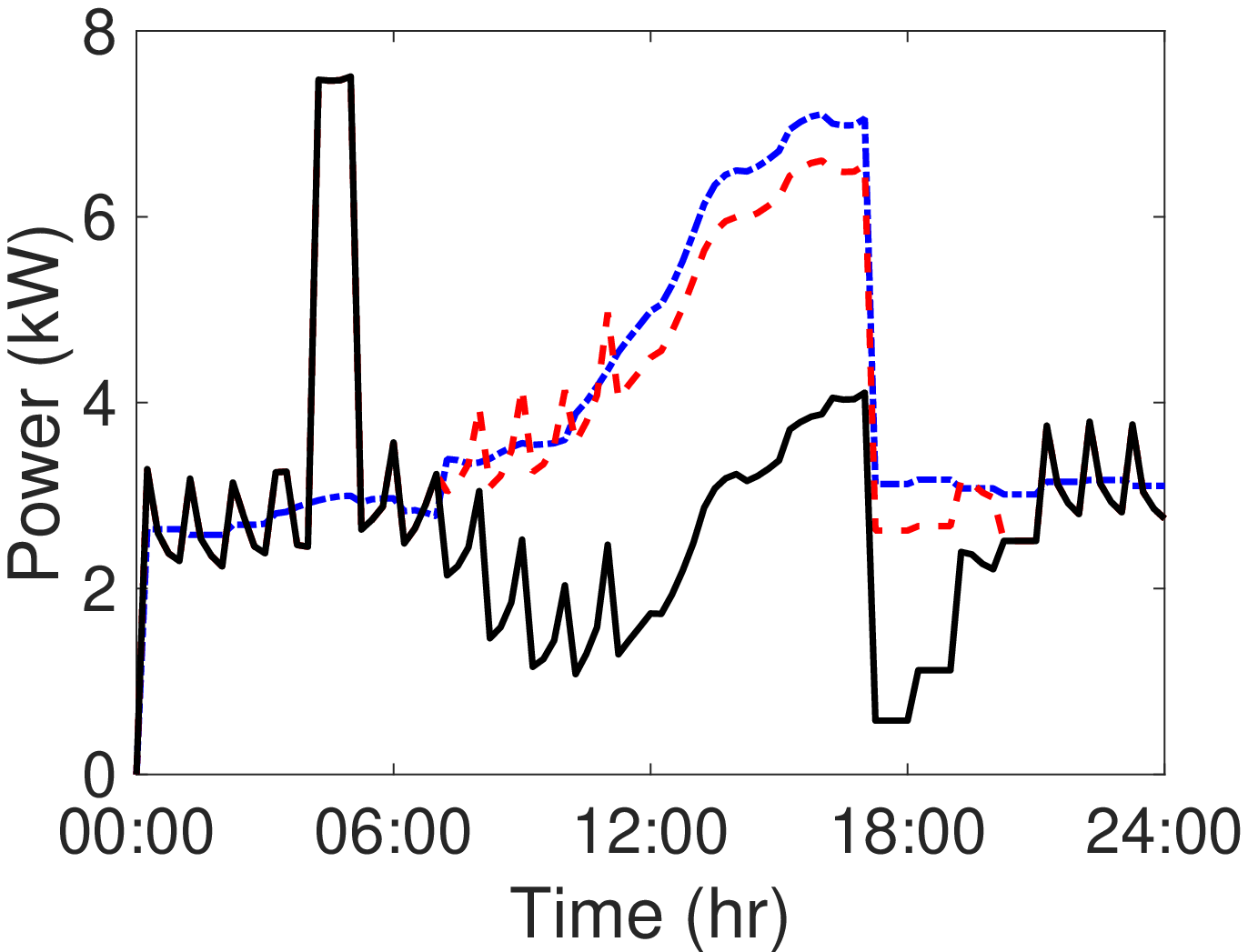}}
		\vspace{-0.1cm}
  \caption{Amount of purchased power during a day under different system configurations for defined occupancy patterns}
\label{fig:12}
\vspace{-0.6cm}
\end{figure}
 
 \subsection{Co-schedule HVAC with PV and Battery Storage}
This section demonstrates the effectiveness of NL-EMPC controller in meeting economic objective of minimizing transacted energy cost by coordinating HVAC control with battery energy storage, PV system, and other inflexible loads of the building. The model simultaneously utilizes the occupancy information and TOU prices to optimally schedule all resources. The PV panel is rated at 4 kW. The actual PV generation follows sun's irradiation during the day and is depicted in Fig. \ref{fig:7}. It is assumed that the battery is in the minimum state of charge ($SOC=0.25$) at the beginning of one-day simulation (00:00). Fig.\ref{fig:8} shows the demand profile for other essential and inflexible loads in the building for a day; the demand varies based on the building occupancy patterns. 

Fig. \ref{fig:12}, shows the amount of purchased power during each hour of the day to satisfy the power requirement of building's HVAC system and inflexible loads for both occupancy patterns: occupancy pattern 1 (Fig. \ref{aaa}) and occupancy pattern 2 (Fig. \ref{bbb}). Three cases are demonstrated: (1) without battery storage and PV, (2) with battery storage but without PV, and (3) with both battery storage and PV. As it can be observed in both sub-figures, with the help of PV, a reduction in total amount of purchased power from the grid is observed. This reduction reaches to it's maximum at around 12:00 noon when there is maximum solar radiation. Similarly, the positive effect of battery can be observed; HEMS decides to purchase the energy at times that TOU electricity tariffs are low (00:00- 07:00) to charge the battery and uses it at the time that TOU electricity tariffs are high (17:00-21:00).    

Table \ref{my-label} shows the cost of electricity for each combination of resources to be co-scheduled. As it can be observed, NL-EMPC can effectively co-schedule all resources and results in significant reducing in the net cost electricity usage. Notice that on co-scheduling HVAC with battery, the cost of electricity usage is decreased for the day. As expected, co-scheduling HVAC with both PV and battery storage leads to the most savings in electricity usage cost. 


\begin{table}[t]
\centering
\vspace{-0.1cm}
\caption{Total cost of electricity usage during a day (24-hours)}
\vspace{-0.3cm}
\label{my-label}
\begin{tabular}{|l|c|c|c|}
\cline{1-4}
 \multicolumn{4}{|c|}{System Configurations} \\ \cline{1-4} 
 & Loads & {Loads+Battery} & {Loads+Battery+PV} \\
\hline
\multicolumn{1}{|c|}{Occupancy 1} & \$3.54 & \$3.38 & \$1.97 \\ \hline
\multicolumn{1}{|c|}{Occupancy 2} & \$3.98 & \$3.82 & \$2.41 \\ \hline
\end{tabular}%
\vspace{-0.5cm}
\end{table}

\section{Conclusion}
\label{Conclusion}
In this paper, we present a NL-EMPC to co-schedule building's HVAC system with its inflexible loads, PV system and battery storage. The proposed NL-EMPC controller is able to optimize the building's electricity usage cost by leveraging the known building's occupancy information while considering an imperfect prediction of the disturbance for HVAC system. The simulation results demonstrate that the proposed controller leads to a reduction in the net-cost of electricity usage while satisfying building occupants' comfort-level. 
\ifCLASSOPTIONcaptionsoff
  \newpage
\fi
	\bibliographystyle{ieeetr}
	
	\bibliography{references}
\end{document}